\documentclass{iopart}

\usepackage{graphicx}

\def\la{\hbox{{\lower -2.5pt\hbox{$<$}}\hskip -8pt\raise
-2.5pt\hbox{$\sim$}}}
\def\ga{\hbox{{\lower -2.5pt\hbox{$>$}}\hskip -8pt\raise
-2.5pt\hbox{$\sim$}}}

\begin{document}

\begin{center}\title{The 2pt+: an enhanced 2 point correlation function}\end{center}

\author{M. Ave$^{1,2}$, L. Cazon$^{1}$, J. Cronin$^{1}$, J. R. T. de Mello Neto$^{1,3}$, A. V.\ Olinto$^{1,4}$, V. Pavlidou$^{1,5,6}$, P. Privitera$^{1}$, B. B. Siffert$^{1,3,7}$, F. Schmidt$^{1}$, T. Venters$^{1}$}
\address{$^1$ Department of Astronomy and Astrophysics, Kavli Institute for Cosmological Physics, Enrico Fermi Institute, The University of Chicago, 5640 S. Ellis, Chicago, IL 60637, USA}
\address{$^2$ Excellence Cluster Universe, Technische Universit\"at M\"unchen, Boltzmannstr. 2, D-85748, Garching, Germany} 
\address{$^3$  Universidade Federal do Rio de Janeiro, Instituto de F\'isica, Rio de Janeiro, RJ, Brazil}
\address{$^4$ Laboratoire Astroparticule et Cosmologie (APC), Universit\'e Paris 7/CNRS,  10 rue A. Domon et L. Duquet, 75205 Paris Cedex 13, France}
\address{$^5$ Astronomy Department, California Institute of Technology, Pasadena, CA 91125, USA}
\address{$^6$ Einstein (GLAST) Fellow}
\address{$^7$ Universit\`a degli Studi di Napoli Federico II, Via Cinthia, 80126, Napoli, Italy}

 \date{\today}

\begin{abstract}
We introduce a new method for testing departure from isotropy of points on a sphere based on  an enhanced form of the two-point correlation function that we named 2pt+. This method uses information from the two extra variables that define the vector between two points on a sphere. We show that this is a powerful method to test departure from isotropy of a distribution of points on a sphere especially when the number of events is small. We apply the method to a few examples in astronomy and discuss the relevance for limited datasets, such as the case of ultra-high energy cosmic rays.
\end{abstract}

%\pacs{PACS: }
\maketitle

\section{Introduction}

Determining the degree of anisotropy in the sky distribution of astronomical objects, background radiation fields, and transient events is often key in unveiling the hidden nature of an astronomical phenomenon. A number of tests for particular anisotropic patterns are available in the literature such as 
spherical harmonic decomposition, tests for specific angular scale fluctuations (multipole-specific tests), 2pt correlation function, high-order correlation functions, and so forth. Different types of tests are more useful depending on the character of the anisotropy. Some areas of astronomical study are characterized by small fluctuations on a largely isotropic background (such as the cosmic microwave background, e.g., \cite{1991AdSpR..11..193S}; the large-scale structure of galaxy distributions, e.g., \cite{1952PASP...64..247Z,1971PASP...83..113D}; neutrino astronomy, e.g., \cite{Braun:2008bg}, and the extragalactic GeV gamma-ray background, e.g., \cite{1998ApJ...494..523S}), while other areas see deviations from isotropy with very sparsely populated datasets (such as the sky distribution of pulsars, e.g., \cite{1978MNRAS.185..409M}; gamma-ray bursts before the Compton Gamma-ray Observatory, e.g., \cite{1985MNRAS.212..545S,1980PAZh....6..609M}; and the current data on trans-GZK ultra-high energy cosmic rays, e.g., \cite{Abraham:2007bb,Abraham:2007si}). Here we focus on a new method designed for the sparsely sampled type of datasets.
This case is often challenging as with small datasets false positives are easier to encounter, and guarding against them by increasing the test significance when a signal is treated as real or at least interesting reduces the power of the tests (an extreme example: the single-point dataset is maximally anisotropic, yet fully consistent with isotropy!). 

One way that has been used to overcome these difficulties is to use insight and guidance from theoretical expectations to define appropriate tests for very specific types of anisotropy expected out of a dataset, for example tests that look for distributions of datapoints that follow the distribution of known catalogs (e.g., \cite{Abraham:2007bb,Abraham:2007si,Kashti:2008bw,Ghisellini:2008gb,George2008}). However, one would like to define a test that would be ``democratic'' in its treatment of angular scales, ``free'' of any underlying assumption about the type of anisotropy, and as powerful as possible. Clearly not all requirements above can be maximized at the same time, so what we propose here is a reasonable compromise.

In this work we present a statistical test which aims to quantify, in a manner as general as possible, how (in)frequently a particular realization of the positions of $N$ points on a sphere arises from isotropy (i.e., from an underlying spatial distribution with a flat probability density function). 
The outcome of the proposed test for a particular ``observed'' data set will be a significance\footnote{In the strict sense of a classical statistics test, the quantity we are calculating is a {\em p-value}.},
expressing the fraction of datasets of $N$ points drawn from an isotropic
distribution that returns a value for the estimator (test statistic)
which is equally or more rare than the  ``observed' one, in the limit that the experiment is repeated $\rightarrow \infty$ times.
If this fraction is very low, then the observer can claim either that the ``observed'' dataset is a very rare realization of isotropy, or that the underlying {\it ansatz} of isotropy is not valid. 
 
For the purposes of this test, we {\em quantitatively} model isotropy on a sphere as follows: when $N$ points are drawn from an isotropic distribution on a sphere, then the probability\footnote{Here defined as the fraction of positive outcomes in the limit of infinite repetitions of the experiment, i.e., infinite number of draws of $N$ points.} that $n$ points out of the total $N$ lie within the boundaries of an arbitrary surface ($S_0$) follows the binomial distribution:
\begin{equation}
p_{n,N}=C_{N}^{n} p^n(1-p)^{N-n} \,
\label{binomial}
\end{equation}
where $p$=$\frac{S_0}{S_{Tot}}$, and $S_{Tot}$ is the surface of the sphere (4$\pi$ for the
 sphere of radius 1). 

Our aim is to build a test that is comparably sensitive to fluctuations on a wide range of angular scales, for $N$ as low as $20$ up to a maximum value limited by the available computing power (currently, computations for $N$ as high as $\sim 100$ are accessible using a desktop computer).  To check the sensitivity of the test we use mock  anisotropic realizations sampled from a {\em small} number of equal flux point sources  distributed isotropically  and spherical harmonics with
 {\em low}  multipole moment, l. The test is expected to be sensitive in the case of mock maps drawn from a small number of point sources (which have power on small angular scales) and also for distributions with low multipole moments (which have power on large angular scales).
These two cases cover the two extremes of fluctuations when classified in angular scale.

\begin{figure}[htbp]
\begin{center}
\includegraphics[width=5.5in]{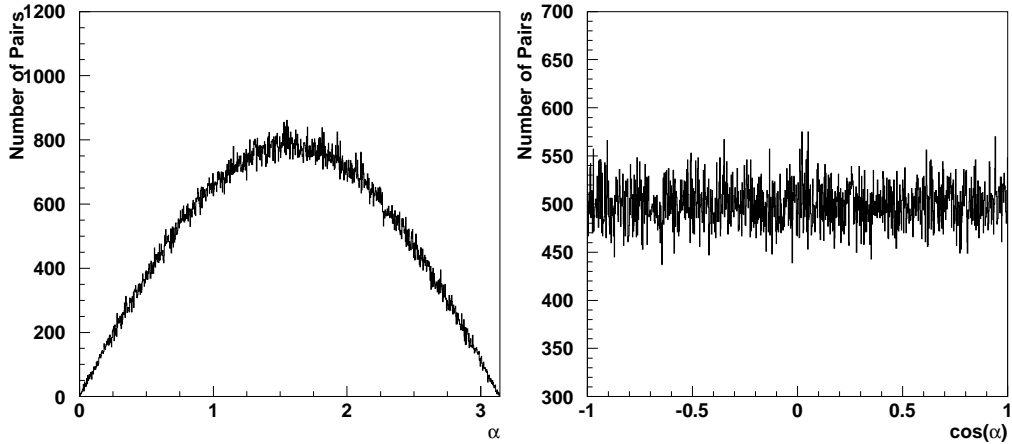}  
\caption{The distribution of $\alpha$ (left panel) and $\cos \alpha$ (right panel) in 1000 realizations of 50 events (1225 pairs) drawn from isotropy. The values of $\cos \alpha$ are consistent with originating from a uniform distribution. }
\label{Alpha}
\end{center}
\end{figure}

Another commonly used statistical test for small datasets is based on the two-point (2pt) correlation function. For this test, the angular distances between pairs of points are calculated, and some measure of the deviation of their distribution from what is expected from isotropy is used as an estimator. The distribution of angular distances between 1225 pairs ($N=50$)  drawn from an isotropic distribution (test repeated for 1000 realizations) is shown in the left panel of 
 Fig. \ref{Alpha}. Here, $\alpha$ is the angular distance between two
 points in rad, while the number of pairs for a data set of $N$ events is $\frac{N  (N-1)}{2}$.  
Higher correlation functions can also be calculated, but it is not straightforward  to extract the independent information contained in these higher order terms. Instead, here we  build upon the 2pt test, by considering two new variables in addition to $\alpha$, which are also related to pairs of points but are {\em uncorrelated} with $\alpha$. The additional information used in this 2pt test motivates the name 2pt+.
 
\section{Basic concept of the 2pt+ test}

\begin{figure}[htbp]
\begin{center}
\includegraphics[width=5.5in]{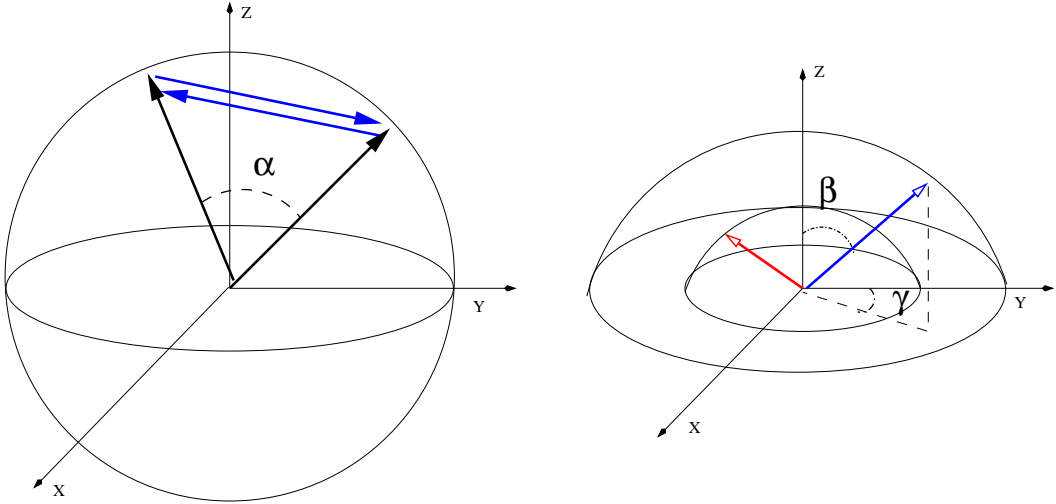}  
\caption{Schematic depiction of the variables used in the 2pt+ test. Left panel: angular distance $\alpha$ between two events. Here, the drawn sphere corresponds to the observed celestial sphere. The black vectors correspond to vectors extending from the origin (observer) to each of the two events. The blue vectors represent the two choices for the definition of the vector {\em between} events. In this work, we always use the vector with a positive $z-$component (in the case depicted here, the vector pointing from right to left). $\alpha$ is the angle subtended by the two event vectors. Right panel: angles $\beta$ and $\gamma$. Here, the vectors {\em between} events have been transported to the origin, and each of the two vectors corresponds to a different pair between events. The length of each vector depends on $\alpha$, so the radii of the two spheres drawn here quantify the angular distance between each pair of events. The angles $\beta$ and $\gamma$ for one of the pairs (the one represented by the blue vector) are shown in the figure. }
\label{CoordinateScheme}
\end{center}
\end{figure}

\begin{figure}[htbp]
\begin{center}
\includegraphics[width=5.5in]{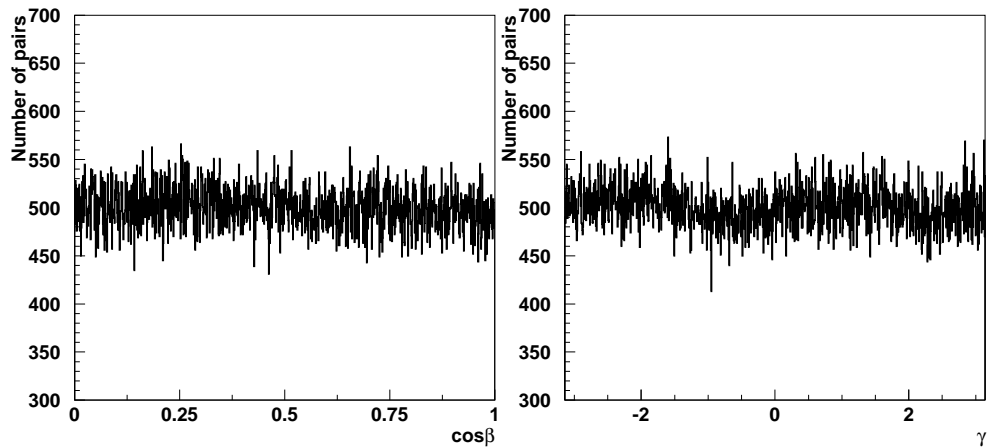}  
\caption{ The distribution of  $\cos \beta$ (left panel) and $\gamma$ (right panel) in 1000 realizations of 50 events (1225 pairs) drawn from isotropy. The values of both quantities are consistent with originating from uniform distributions.}
\label{BetaGamma}
\end{center}
\end{figure}

Fig. \ref{CoordinateScheme} schematically depicts the definitions of the variables that we use in this work. The left panel shows the vector between two points on a sphere, subtending an angle $\alpha$, which we aim to describe using three independent variables. There is an intrinsic degeneracy in the definition of this vector, as either one of the points in the pair can be regarded as the vector origin. In this work, we always choose the point of origin so that the $z-$component of the vector is positive. The vector between  two events  translated to the origin of coordinates, can be described using the following 3 variables as in the right panel of Fig. \ref{CoordinateScheme}: 
\begin{enumerate}
\item $\cos \alpha$, which is a measure of the length of the vector;
\item $\cos \beta$, which is the cosine of the vector's polar angle; and
\item $\gamma$, which is the vector's azimuthal angle. 
\end{enumerate}
The distribution of the values of these variables for 1000 realizations of 50 events and 1225 pairs drawn from an underlying isotropic spatial distribution is shown in the right panel of Fig.~\ref{Alpha} and in Fig.~\ref{BetaGamma}. These plots show that  the frequency distribution of the three variables is consistent, to the accuracy of our testing,  with being uniform  (equal fraction of draws for all bin values).
The two new variables introduced here in addition to the angular distance between pairs ($\cos \beta$,$\gamma$) are sensitive to the orientation of the pairs and therefore to features
 in the sky that have preferential directions, such as a plane in the sky, a network of overdense regions, and filamentary structures.

\begin{figure}[htbp]
\begin{center}
\includegraphics[width=5.5in]{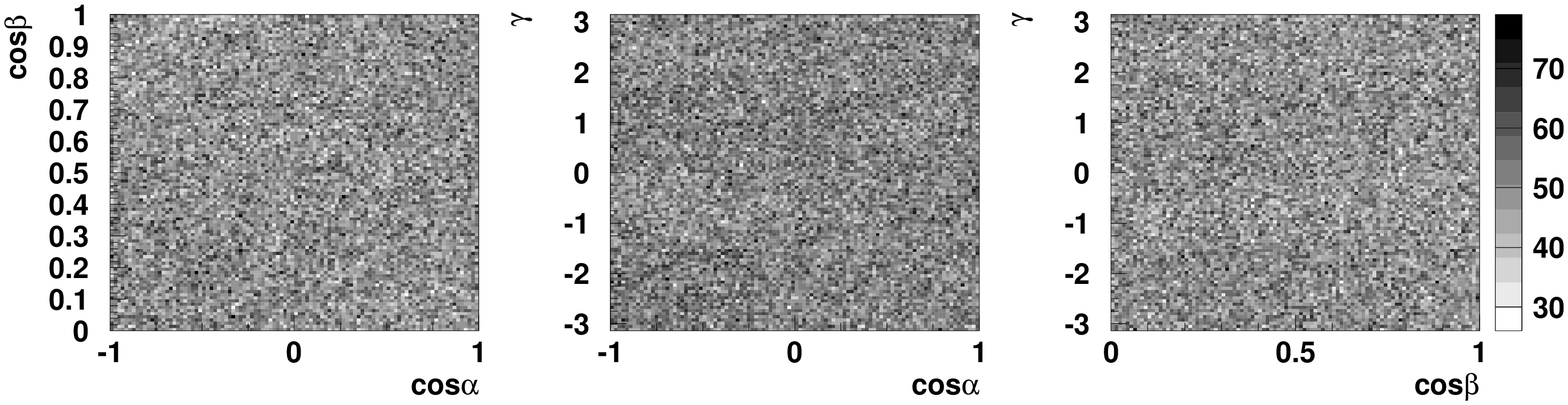}  
\caption{Histograms of $\cos \alpha$-$\cos \beta$, $\cos \alpha$-$\gamma$ and $\cos \beta$-$\gamma$ values for draws from isotropy. No apparent pattern or correlation between any of the variable pairs is seen. }
\label{Corr}
\end{center}
\end{figure}

Fig. \ref{Corr} shows two-dimensional histograms of the values of our three variables in pairs of two, for draws from isotropy. 
In this way, we visualize any potential correlation between each pair of our three variables, when averaged over many realizations, which is clearly very small.  The lack of correlation between $\cos \alpha$ and our two additional variables, $\cos \beta$ and $\gamma$, 
 implies that the two new variables encode independent information that  we add to the power of the standard two-point correlation function by incorporating them into our estimator. In addition, the lack of correlation between the two variables indicates that each brings to the test independent pieces of information, and the power of the test is maximized by incorporating both.

However, it should be noted that there is a major difference between $\cos \alpha$ and  ($\cos \beta$,$\gamma$):
 the set of $\cos \alpha$ values for a data set is independent of the reference system in which it is calculated, while
 the values of  ($\cos \beta$,$\gamma$) are not.  Any reference system on the sphere is related to each other through a rotation and
 any statistical test should be invariant under rotations. If we consider a particular data set in the plane
  ($\cos \beta$,$\gamma$) (right panel in Fig. \ref{Corr}), a rotation is a one-to-one mapping, and neighbouring points  remain neighbouring (and their angular distances  remain unchanged) after the transformation. Therefore, to minimize any possible dependence on rotations we  adopt the following strategy: 
First, a  binned likelihood test is applied to assess the flatness of the $\cos \alpha$ distribution, and  a significance  ($S_\alpha$) is obtained from it.
Then, a second binned likelihood test is applied to assess the flatness of the distribution in the ($\cos  \beta$,$\gamma$) plane, obtaining a significance $S_{\beta\gamma}$.
The two significances are combined using Fisher's method. Any remaining dependence on rotations will be quantified below.

\section{Implementation of the 2pt+ test}

To calculate $S_\alpha$ and $S_{\beta\gamma}$, binned likelihood
estimators are used. A data set with $N$ events is a collection of
$\frac{N   (N-1)}{2}$ values of $\cos \alpha$ and ($\cos 
\beta$,$\gamma$). Since the distributions are flat, if we divide the
phase space into equal parts we expect on average the same number of
events in each one. In the case of $\cos \alpha$ we divide the phase space (-1,1)
into equal parts in such a way that the expected number of events for
each bin is $\mu$ (thus, the number of bins $N_{bins,\alpha}$ is the integer closest to $\frac{N (N-1)}{2  \mu}$). A similar approach is used for the 
($\cos \beta$,$\gamma$) plane, but since this is a two dimensional
distribution, the number of bins in each axis  is  the integer closest to $N_{bins,\beta\gamma}=\sqrt { \frac{N  (N-1)}{2  \mu} } $.

The two pseudo-likelihood estimators are defined as follows:
\begin{eqnarray}
 L_\alpha &=&\prod_{i=0}^{N_{bins,\alpha}} P(n^{\alpha,obs}_{i},\mu) \,  \nonumber \\
 L_{\beta\gamma} &=& \prod_{j,k=0}^{N_{bins,\beta\gamma}} P(n^{\beta\gamma,obs}_{jk},\mu) \,
\label{EqnEstim}
 \end{eqnarray}
where $P(n^{\alpha,obs}_{i},\mu)$ ($P(n^{\beta\gamma,obs}_{jk},\mu)$) is the Poisson distribution with mean $\mu$
 and $n^{\alpha,obs}_{i}$ ($n^{\beta\gamma,obs}_{jk}$) observed number of pairs in the $i^{th}$  $\cos \alpha$ bin ($j^{th}k^{th}$ ($\cos 
\beta$,$\gamma$) bin). We have verified through Monte Carlo simulations that the distribution in each bin is sufficiently similar to Poisson. 

\begin{figure}[htbp]
\begin{center}
\includegraphics[width=3.5in]{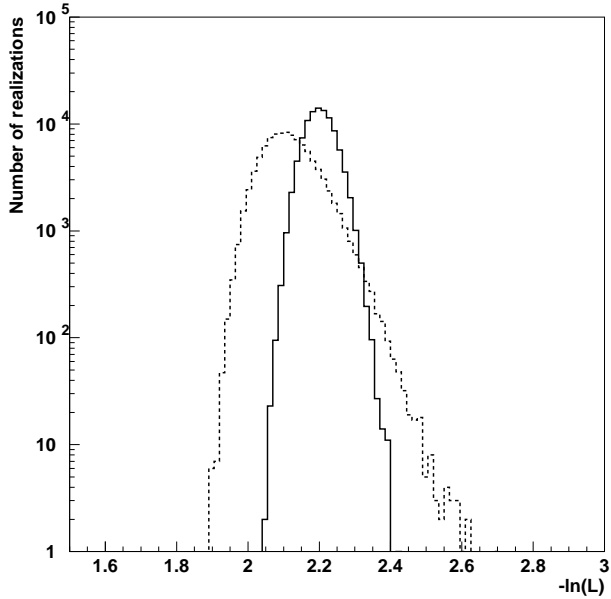}
\caption{The distribution of the estimators $L_\alpha$ (solid line) and $L_{\beta\gamma}$(dotted line), for event draws from an isotropic distribution. If the two estimators were truly independent, the distributions would be identical. Small correlations are introduced by considering pairs instead of events on the sphere, causing the deviation of their Fisher-combined significance from a random variable uniformly distributed between 0 and 1, as seen in Fig. \ref{iso-sgf}. 
}
\label{iso-estimators}
\end{center}
\end{figure}

\begin{figure}[htbp]
\begin{center}
\includegraphics[width=5.5in]{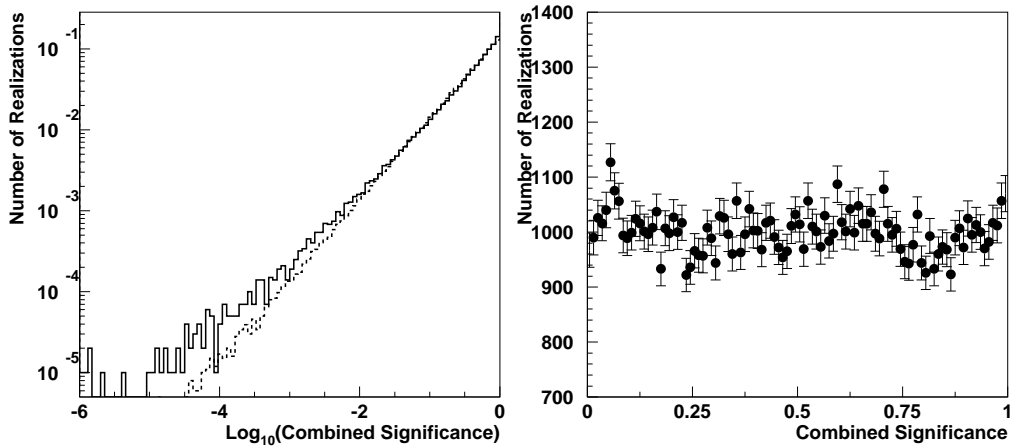}
\caption{Left panel: combined significances obtained with Eq. \ref{SignificanceCombination} (solid line) compared with the expected distribution if the
 two estimators are completely independent (dotted line). Right panel: the distribution of $S^{\rm corr}_{\rm combined}$ for 50 events and $\mu$ = 5.}
\label{iso-sgf}
\end{center}
\end{figure}

Fig. \ref{iso-estimators} shows the distribution of the two estimators for an isotropic distribution with $N$=50 and $\mu$=5. It should be noted that the number of factors entering the calculation of $L_\alpha$ and $L_{\beta\gamma}$ (see  Eq. \ref{EqnEstim}) are the same. Therefore, if  each one is completely independent from each other, the distribution of the two estimators should be the same. This is not the case, reflecting correlations introduced when considering event pairs as the observable instead of  the individual points on the sphere. The small correlations can be easily corrected for as we discuss below (see Fig. \ref{iso-sgf}).

To calculate the significance for a given realization, we integrate the normalized distributions in Fig. \ref{iso-estimators}, obtaining  $S_\alpha$ and $S_{\beta\gamma}$. If these significances quantify {\em truly independent experiments}, the combined 
 significance can be calculated analytically using 
Fisher's method. The method suggests that if 2 independent tests,  each yielding a significance (p-value), $S_\alpha$ and $S_{\beta \gamma}$,  are combined, then the variable
$-2 (\ln S_\alpha + \ln S_{\beta \gamma})$
is distributed as chi-square with $4$ degrees of freedom, and the combined significance can be obtained by integrating under the tail of this distribution above its ``observed'' value defined by the two individually measured significances. The probability distribution function (PDF) of the chi-square with 4 degrees of freedom, if the random variable is $Y$, is 
$f(Y, 4) = Y\exp[-Y/2]/4$,
and the significance for an observed value $Y_0$ (i.e., the PDF integral for $Y$values above $Y_0$) is 
\begin{equation}
S_{Y_0} = \int_{Y_0}^{\infty}f(Y,4)dY 
= \left(1+\frac{Y_0}{2}\right)\exp[-Y_0/2]\,.
\end{equation}
Substituting $Y_0 = -2(\ln S_\alpha + \ln S_{\beta \gamma})$ we get 
\begin{equation}
S_{\rm combined} = S_\alpha S_{\beta \gamma} \left(1-\ln S_\alpha S_{\beta \gamma}\right)\,.
\label{SignificanceCombination}
\end{equation}
This is simply the probability that the product of two random variables uniformly distributed between (0,1) is smaller or equal than the observed product.

The left panel in Fig. \ref{iso-sgf} shows the distribution of log$_{10}S_{\rm combined}$ (solid histogram) together with the distribution of log$_{10}u$ (dotted histogram),
 where $u$ is a random number between (0,1). Deviations for significances smaller than 10$^{-4}$ are apparent and caused by  a small correlation
 between $L_\alpha$ and $L_{\beta\gamma}$. The combined significance $S_{\rm combined}$ is then corrected using Monte Carlo calculations. The distribution of the corrected significance, $S^{\rm corr}_{\rm combined}$, is shown in the right panel of Fig. \ref{iso-sgf}, and is indeed consistent with a uniform distribution.

\subsection{The selection of bin size ($\mu$)}

The least straight-forward part in a binned likelihood is setting the value of $\mu$ to be used, i.e., the bin size.  If we have $N$ random numbers distributed between 0 and 1, the maximum number of independent questions to  be asked is $N$. Therefore, the number of bins should  be smaller than $N$, i.e., $\mu$ (the expectation value for points per bin) should be equal to or larger than 1. In the present case, since $N$ is the number of pairs and enhanced correlations are present, $\mu$ should be clearly larger than 1.

A binned likelihood test is only sensitive to fluctuations at scales larger than the bin size. That is the reason why the bin size should be kept as small as possible when there is no {\it a priori} reason setting a lower-limit to its size. However, the sensitivity to fluctuations at increasingly
 larger scales decreases because the information about coherent fluctuations of bins is lost in the binned likelihood  method. A possible approach is to modify the binned likelihood to include the information about correlations between bins. However, in this work we  adopt a more conservative approach: we  keep the bin size as low as possible and consider the
 resulting sensitivity as a {\em lower bound} to what can in principle be achieved.

To select the minimum reasonable $\mu$, we generate mock anisotropic realizations of 50 events sampled from: a) $n_{ps}$ point sources isotropically distributed, and b)
 multipoles with low values of l. Each case is evaluated 5000 times and each time $S_\alpha$, $S_{\beta\gamma}$ and $S^{corr}_{combined}$ are calculated. 
We  use the fraction of realizations with a significance smaller than 10$^{-4}$ (hereafter $F_4$) as a measure of the sensitivity to each case. 

\begin{figure}[htbp]
\begin{center}
\includegraphics[width=4.5in]{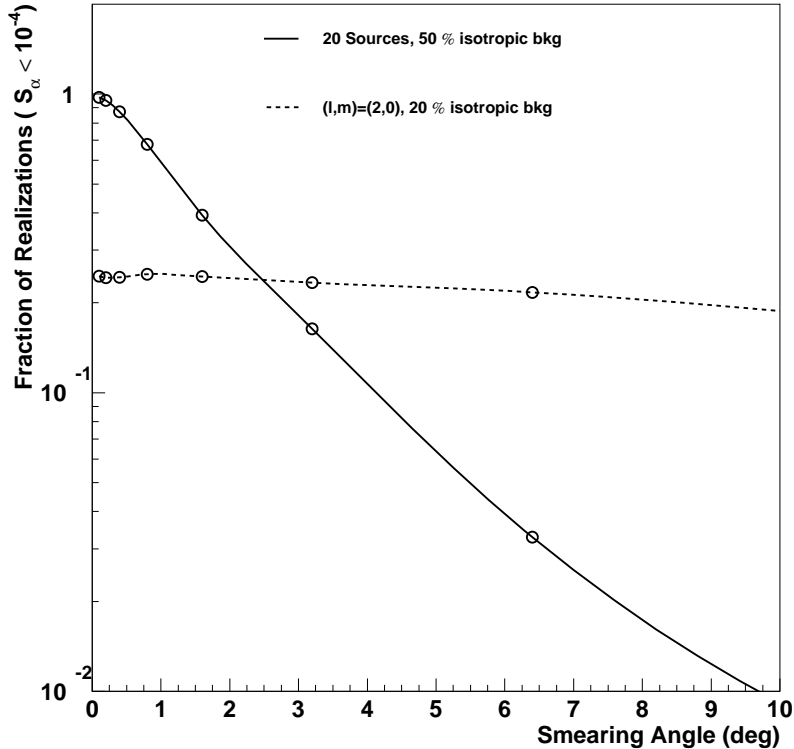}
\caption{Impact of a Gaussian smearing on mock anisotropy maps. Only $L_\alpha$ is used to calculate the significances. Results shown are for $N= 50$  and $\mu=5$. The solid curve is for 25 events from 20 equal flux point sources isotropically distributed over a 50\% isotropic background (i.e., 25 isotropic events). The dashed line is for a dipole with 20\% (i.e., 10 events) of the events from an isotropic background.}
\label{Smearing}
\end{center}
\end{figure}

Fig. \ref{Smearing} shows $F_4$ for two cases with $N= 50$: the solid line is for 20 equal flux point sources distributed isotropically with 50\% isotropic background (i.e., 25 events from point sources and 25 events from the background); while the dashed line shows events sampled from a spherical harmonic, $Y_{\rm l,m}$, with (l,m)=(2,0) with 20\% (10 events) from an isotropic background. Each event is smeared with a Gaussian distribution of variable width indicated in the $x$ axis. This smearing is representative of the angular resolution of the experiment (or physical smearing, e.g., due to lensing). The value of $F_4$ is calculated using only  $S_\alpha$. It is clear  from the multipole case that if the anisotropy manifests itself at angular
 scales larger than the smearing angle, the sensitivity of the test does depend on the smearing. The dependence is strong in the case of point sources. Realizations
of isotropy would correspond to a flat line with $F_4$=$10^{-4}$, i.e., outside of the range shown.  The value of $\mu$ used is 5, showing
 that for this particular value we are sensitive to fluctuations at very different angular scales.

\begin{figure}[htbp]
\begin{center}
\includegraphics[width=5.5in]{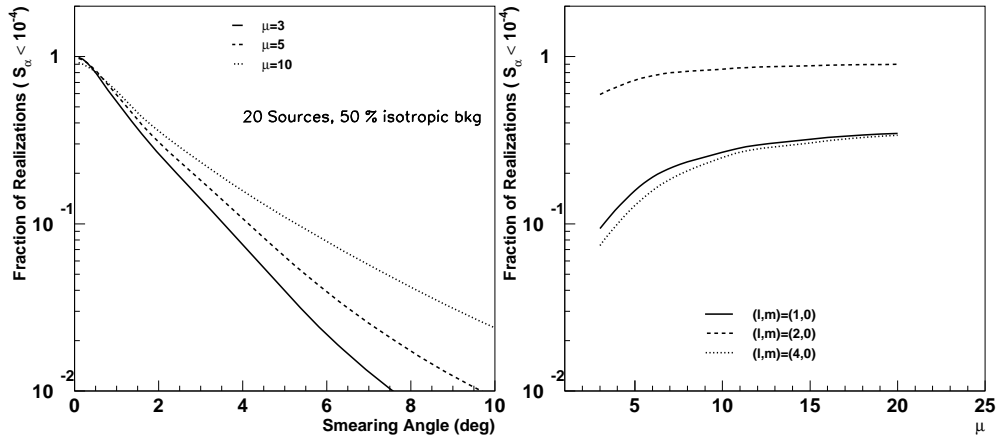}
\caption{The selection of $\mu$ (see text) to calculate $L_\alpha$. Results shown are for $N= 50$. Left panel shows $F_4$ from $S_\alpha$ as a function of smearing angle of 20 point sources for $\mu$= 3 (solid), 5 (dashed), and 10 (dotted). Right panel shows  $F_4$ from $S_\alpha$ as a function of  $\mu$ for spherical harmonics with (l,m)=(1,0) solid, (2,0) dashed, and (4,0) dotted lines.}
\label{Binning1}
\end{center}
\end{figure}

\begin{figure}[htbp]
\begin{center}
\includegraphics[width=4.5in]{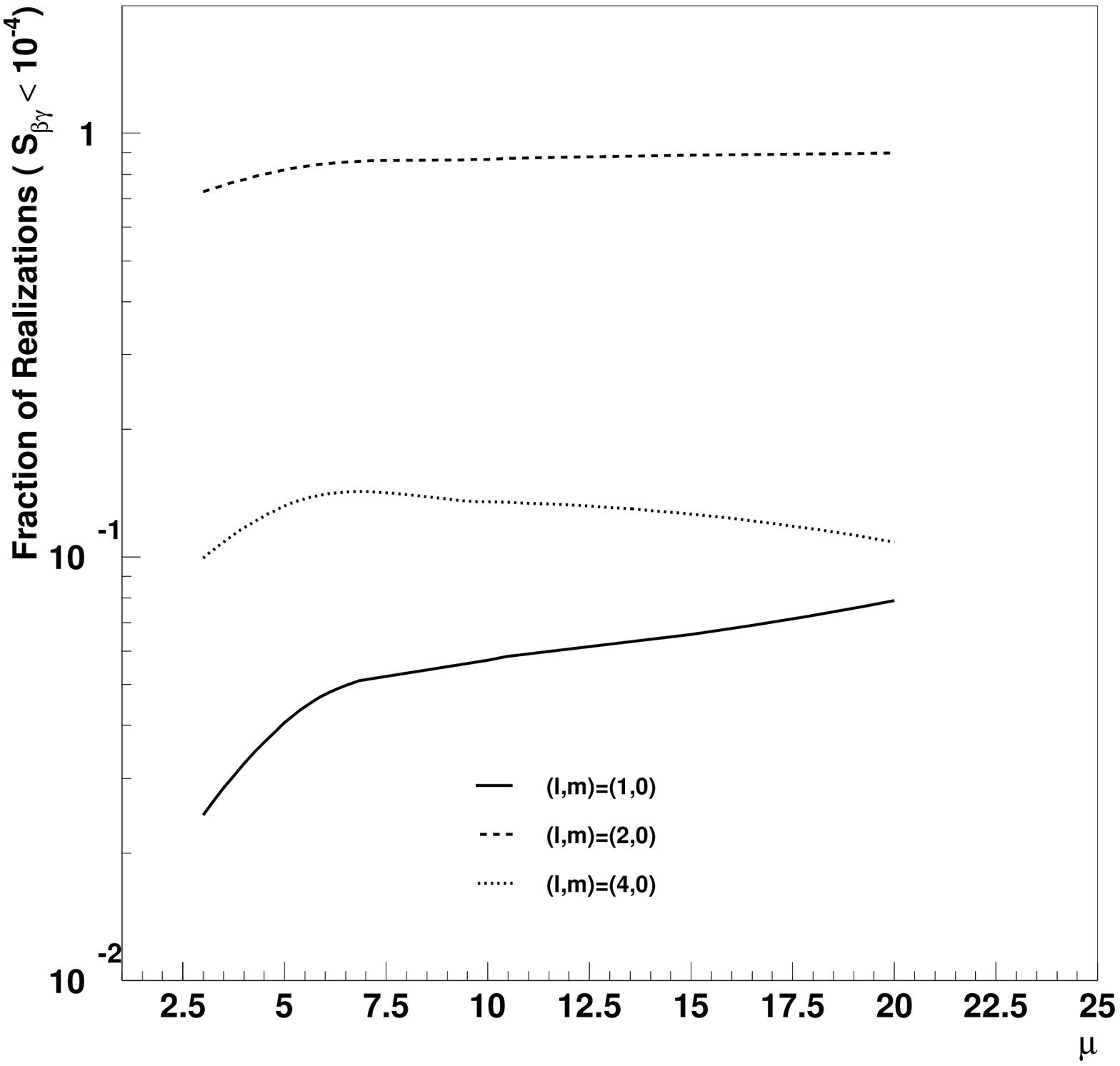}
\caption{$F_4$ from $S_{\beta\gamma}$ as a function of  $\mu$ for spherical harmonics with (l,m)=(1,0) solid, (2,0) dashed, and (4,0) dotted lines. Results shown are for $N= 50$.}
\label{Binning2}
\end{center}
\end{figure}

The left panel in Fig.~\ref{Binning1} shows $F_4$ as a function of the smearing angle for 3 choices of $\mu$ (3, 5, and 10), where only $S_\alpha$ is used to calculate $F_4$.
  The mock samples correspond to 20 point sources and 50\% isotropic background. The trend is clear: for larger smearing angles, larger values of $\mu$ lead to increasing sensitivities.

 The right panel in Fig.~\ref{Binning1} shows $F_4$ calculated with $S_\alpha$ as a function of $\mu$ for realizations sampled from multipoles with different values of (l,m). (In the figure (l,m) is (1,0) for the solid line, (2,0) for the dashed line and (4,0) for the dotted line respectively). The sensitivity increases for values of  $\mu$ between 3 and 10 and is roughly constant after that. 

Fig. \ref{Binning2} shows $F_4$ as a function $\mu$ for realizations sampled from the same multipoles as in the right panel of Fig.\ \ref{Binning1}. This time $S_{\beta\gamma}$ is used to calculate $F_4$. The dependence on  $\mu$ is also mild above 5. 
 
 After this exercise, it is clear that $\mu$=5 is a rather conservative value  to be used either for $\cos \alpha$ or  ($\cos \beta,\gamma$), and it is the one used in this work from now on. We are currently developing an unbinned likelihood estimator that bypasses the problem of selecting $\mu$. The results of this analysis will be reported in a future publication.

\subsection{Impact of a change of coordinates}

\begin{figure}[htbp]
\begin{center}
\includegraphics[width=5.5in]{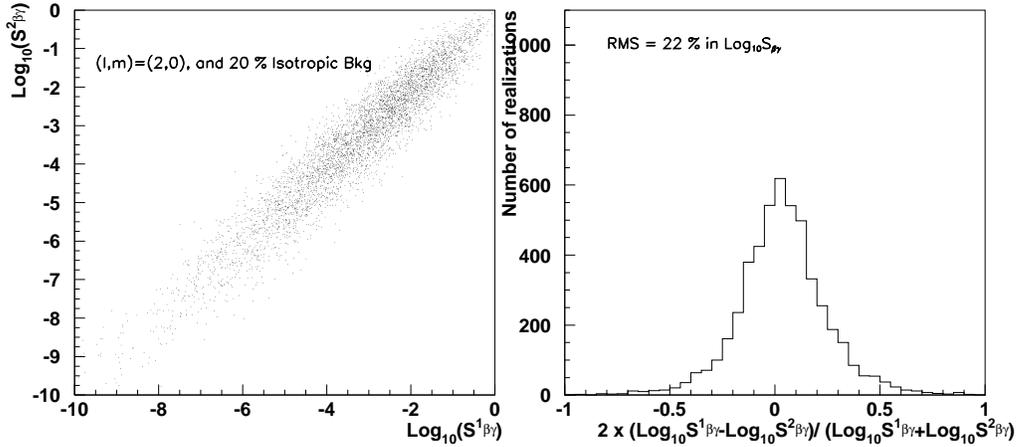}
\caption{Impact of a change of coordinates on the significances calculated using $L_{\beta\gamma}$. Left panel: correlation between $S_{\beta\gamma}$ calculated in each coordinate system. Right panel: distribution of the dispersion of the two significances around the diagonal.}
\label{Rot}
\end{center}
\end{figure}

As discussed above, the set of ($\cos \beta,\gamma$) values depends on the coordinate system used to evaluate them. To check the sensitivity of the test to the coordinate system,  we calculate $S_{\beta\gamma}$ using a reference coordinate system  $S_{\beta\gamma}^1$ and another, randomly chosen coordinate system  $S_{\beta\gamma}^2$. Fig.\ \ref{Rot} shows the result for realizations sampled from a multipole with (l,m)=(2,0). The two significances are not numerically equal, however they show excellent correlation with a spread in Log$_{10}S_{\beta\gamma}$ of $\sim 0.22$. This spread, when propagated to the final significance  $S^{corr}_{combined}$ is further reduced (to $\sim 0.17$). 
So we can conclude that the dependence on the chosen coordinate system in not very significant.
 
\section{The sensitivity of the test}

\begin{figure}[htbp]
\begin{center}
\includegraphics[width=4.5in]{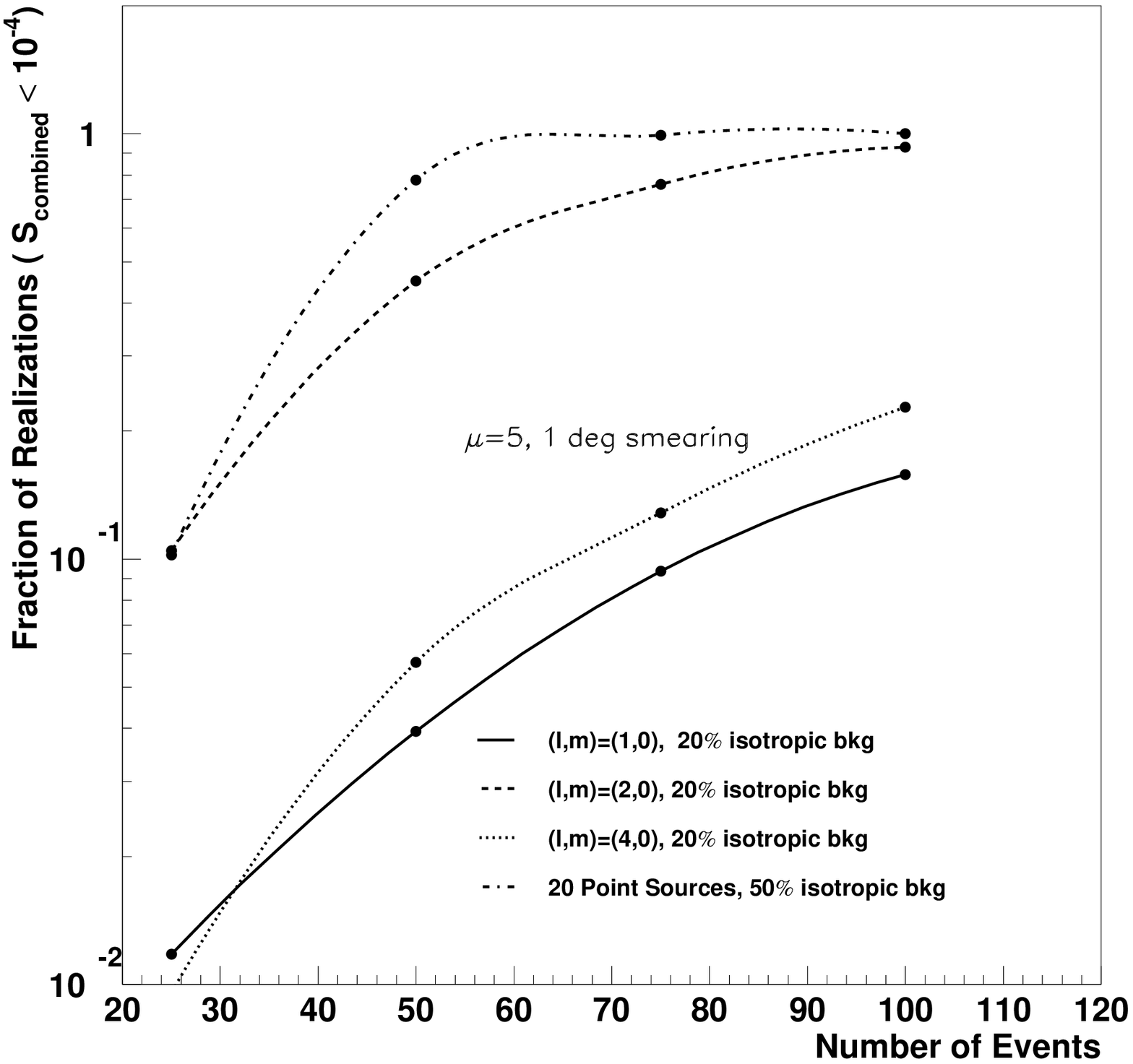}
\caption{Sensitivity of the test as a function of the number of events for 4 cases of anisotropies.}
\label{Dim}
\end{center}
\end{figure}

Fig. \ref{Dim} shows the sensitivity of the test as a function of the number of events in the sphere for 4 anisotropic cases. Three spherical harmonics with 20\% isotropic background are shown ((l,m) = (1,0) solid, (2,0) dashed, (4,0) dotted lines) and the case of 20 point sources with 50\% background is shown in the dot-dashed line.  The evolution of $F_4$ is different for different cases but in all of them there is an increase of sensitivity when the number of events is increased. The test seems to show good sensitivity for number of events between 25 and 100 with anisotropies at rather different angular scales.

\begin{figure}[htbp]
\begin{center}
\includegraphics[width=4.5in]{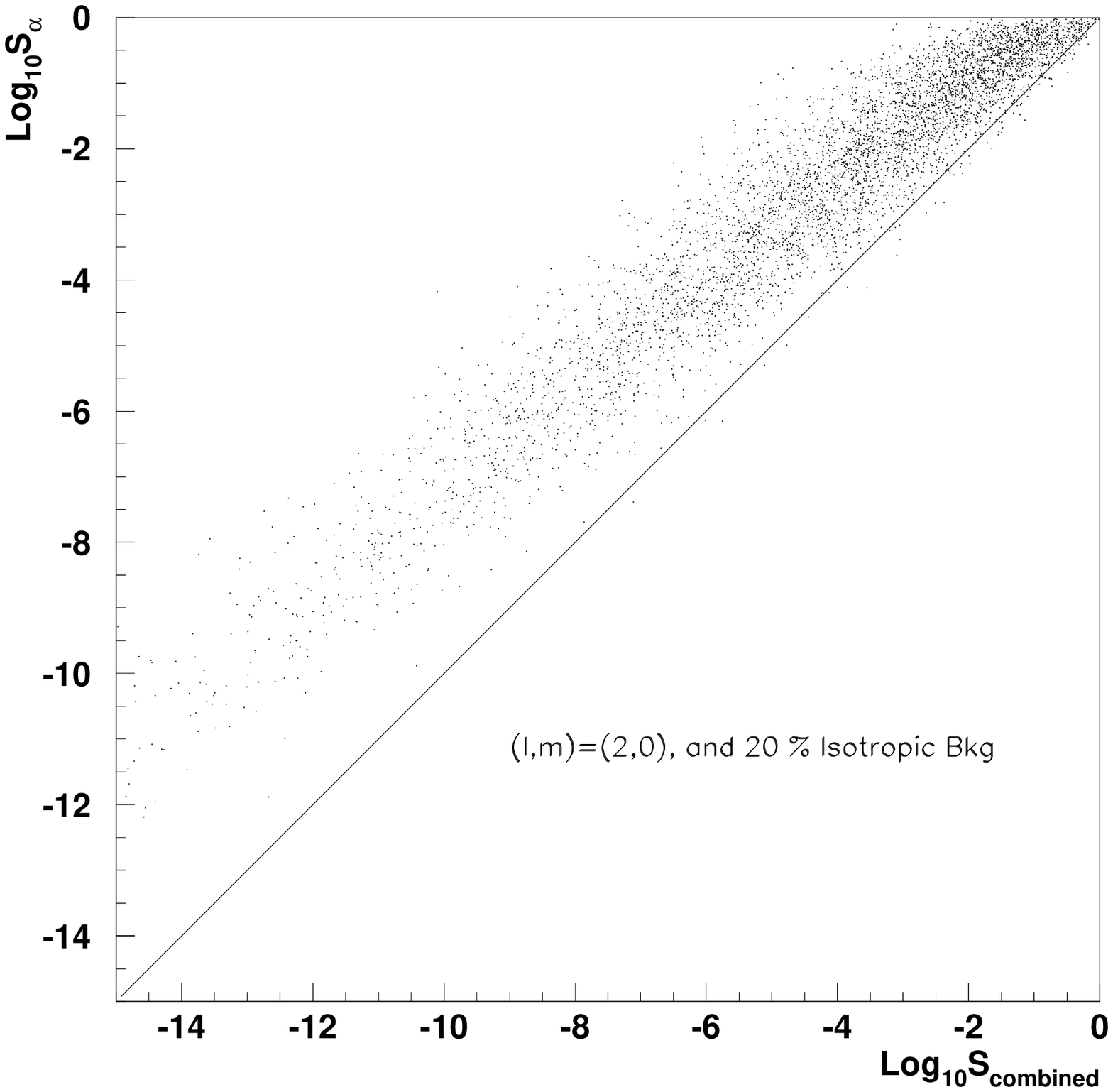}
\caption{Improvement in sensitivity by using the 2pt+ test over the classic 2pt test (here implemented by using $S_\alpha$ alone), for a low-multipole (l=2) anisotropy with a 20\% isotropic background. Each realization corresponds to $N=50$ and $\mu=5$.}
\label{Enhance}
\end{center}
\end{figure}

Fig. \ref{Enhance} demonstrates the sensitivity improvement of the 2pt+ test with respect to the standard 2pt test, which only considers information encoded in $\alpha$. In this figure we logarithmically plot, for a low-multipole underlying distribution (l,m = 2,0) over  a 20\% isotropic background, the significance using information in $\alpha$ alone, $S_\alpha$, as a function of $S^{\rm corr}_{\rm combined}$ for the same realization. The experiment is performed for $N=50$ and with a bin size $\mu=5$.  We see that $S_\alpha$ is in almost all cases higher (worse) than $S_{\rm combined}$. The 2pt+ test is systematically more sensitive than the 2pt test for this type of anisotropy. 

\begin{figure*}[htbp]
\begin{center}
\includegraphics[width=4.5in]{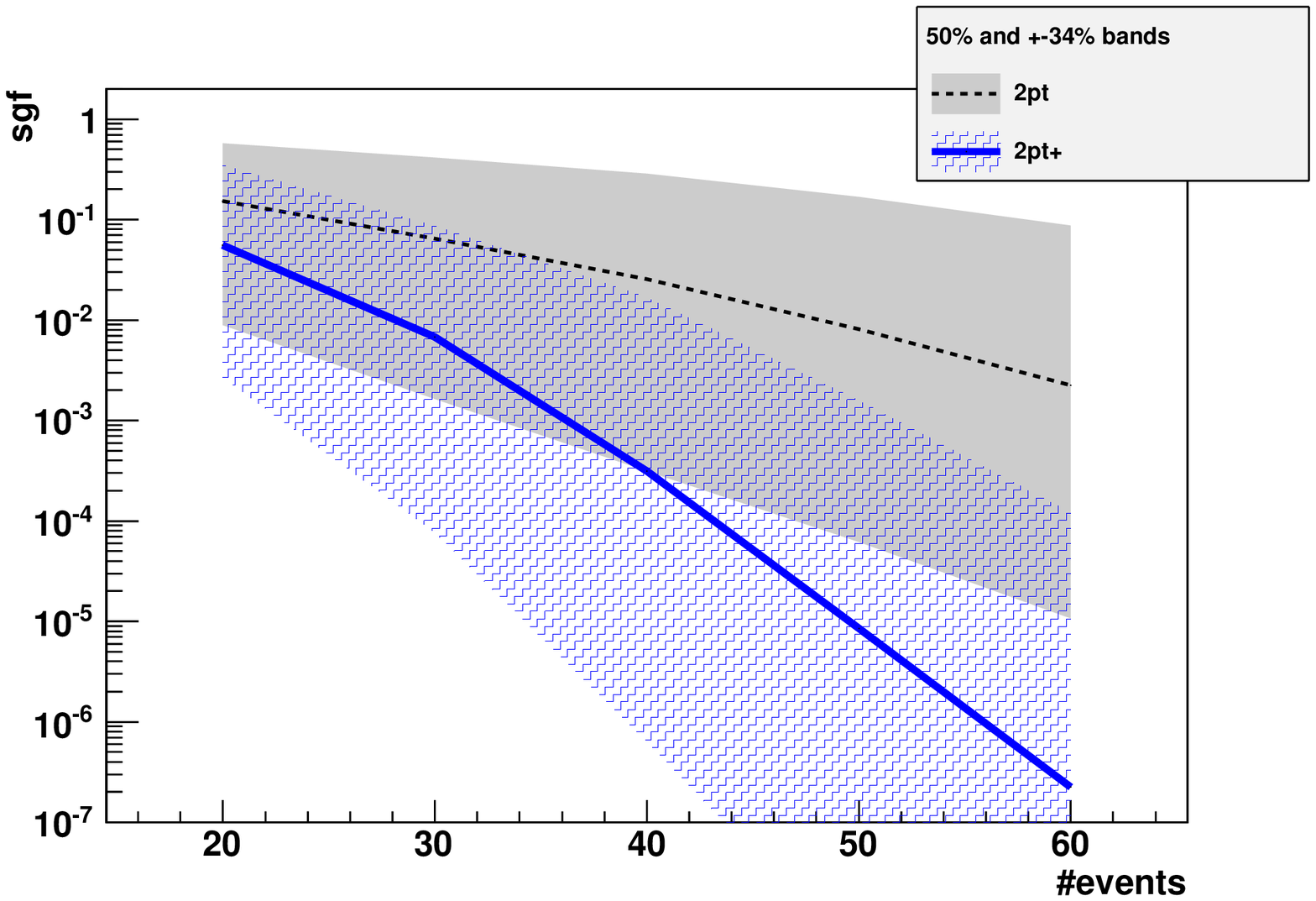}
\includegraphics[width=4.5in]{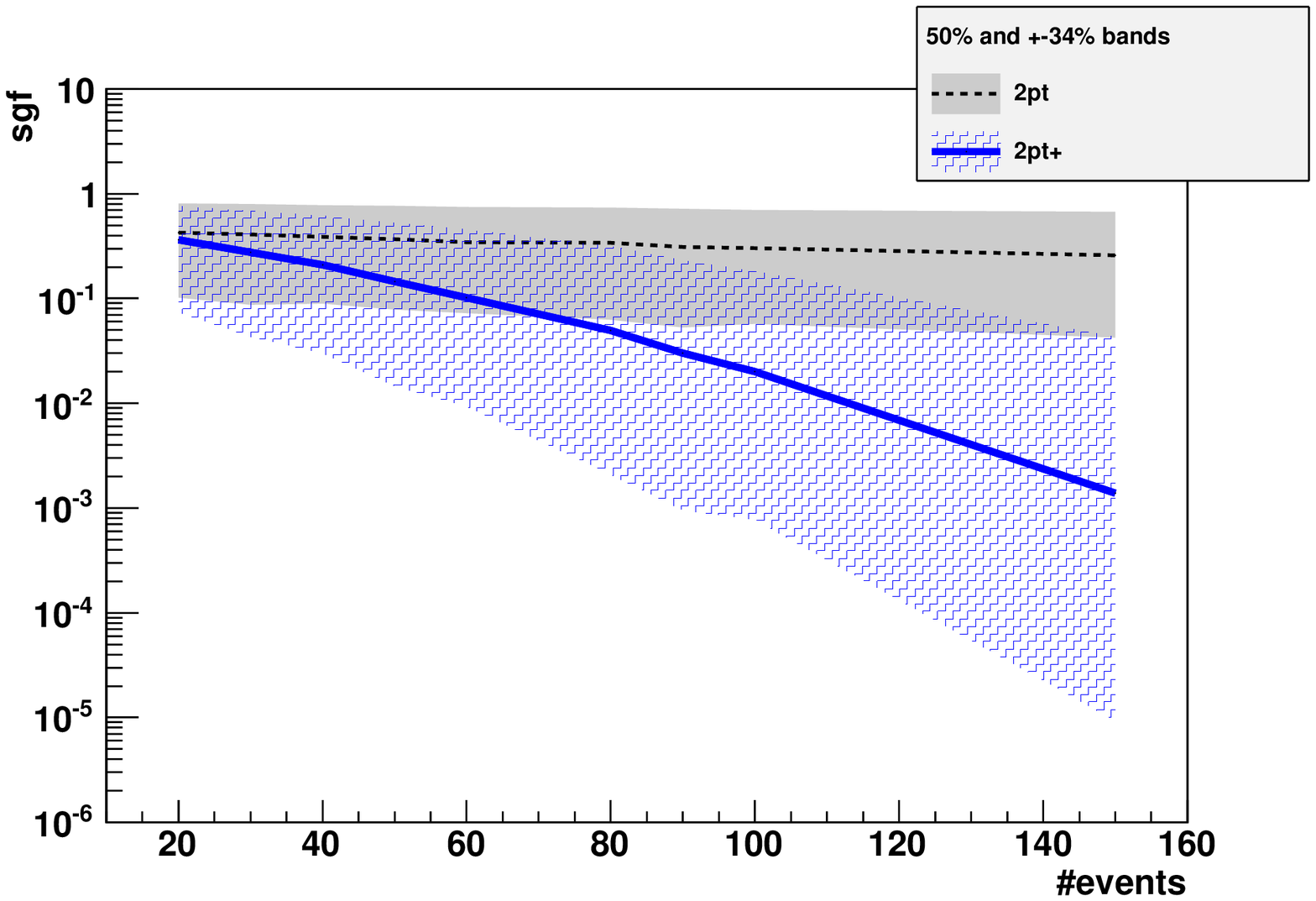}
\caption{ Evolution of the median of the significances and 1 sigma band as a fucntion of the number of events for the 2pt+ and differential 2pt. The events are drawn from the VCV catalogue of sources with $z\leq 0.018$ and 3.2 deg smearing, and taking into account the Auger South exposure. Upper panel has no isotropic background while a 50\% isotropic background is included in the lower panel.}
\label{VCV_sens}
\end{center}
\end{figure*}

Fig. \ref{VCV_sens} similarly demonstrates the sensitivity improvement of the 2pt+ over the standard 2pt test for a type of anisotropy motivated by recent results from the Pierre Auger Observatory on the sky distribution of events with energies above $\sim 6 \ 10^{19}$ eV \cite{Abraham:2007bb,Abraham:2007si}. The distribution of the 27 highest energy events were found to depart from isotropy through a test which shows a correlation of ultra-high energy cosmic ray events with nearby active galactic nuclei (AGN) selected from the  Veron-Cetty \& Veron catalog of quasars and active galactic nuclei (12th Ed.)  \cite{VCV:2006}  (VCV) with redshift  $z\leq 0.018$ and a Gaussian 3.2 degree smearing.  This recent result is one of the most prominent cases of sparsely-sampled datasets that need to be tested for compatibility with an isotropic distribution on the sky, and a natural application for the 2pt+ test.

On the left panel of Fig. \ref{VCV_sens}, the 2pt+ and the classic 2pt (here implemented using $S_\alpha$ alone) are shown for a mock catalog with varying number of events (from 20 to 60) drawn from the VCV catalog of AGN with the same parameters chosen by the Auger results. The right panel shows the same functions for the same catalog plus an isotropic background with 50\% of the events. 
The black and blue solid lines correspond to the median significance of the 2pt and 2pt+ tests as a function of $N$ respectively, while the bands around these lines correspond to the behavior of $\pm 34\%$ of sets around the median. In both cases the test performs better than the classic 2pt test. The relative improvement {\em increases } with increasing number of events. In addition, in the case where the anisotropic signal is added to a 50\% isotropic background (lower panel), the median significance of the 2pt+ test reaches sub-percent levels above 100 events, while hardly any overall improvement in significance is seen with the 2pt test: for the case of a mild signal over a substantial isotropic background, the 2pt+ test can pick up the deviation from isotropy with samples so sparse that the 2pt test would show no sensitivity whatsoever. 

\begin{figure}[htbp]
\begin{center}
\includegraphics[width=3in]{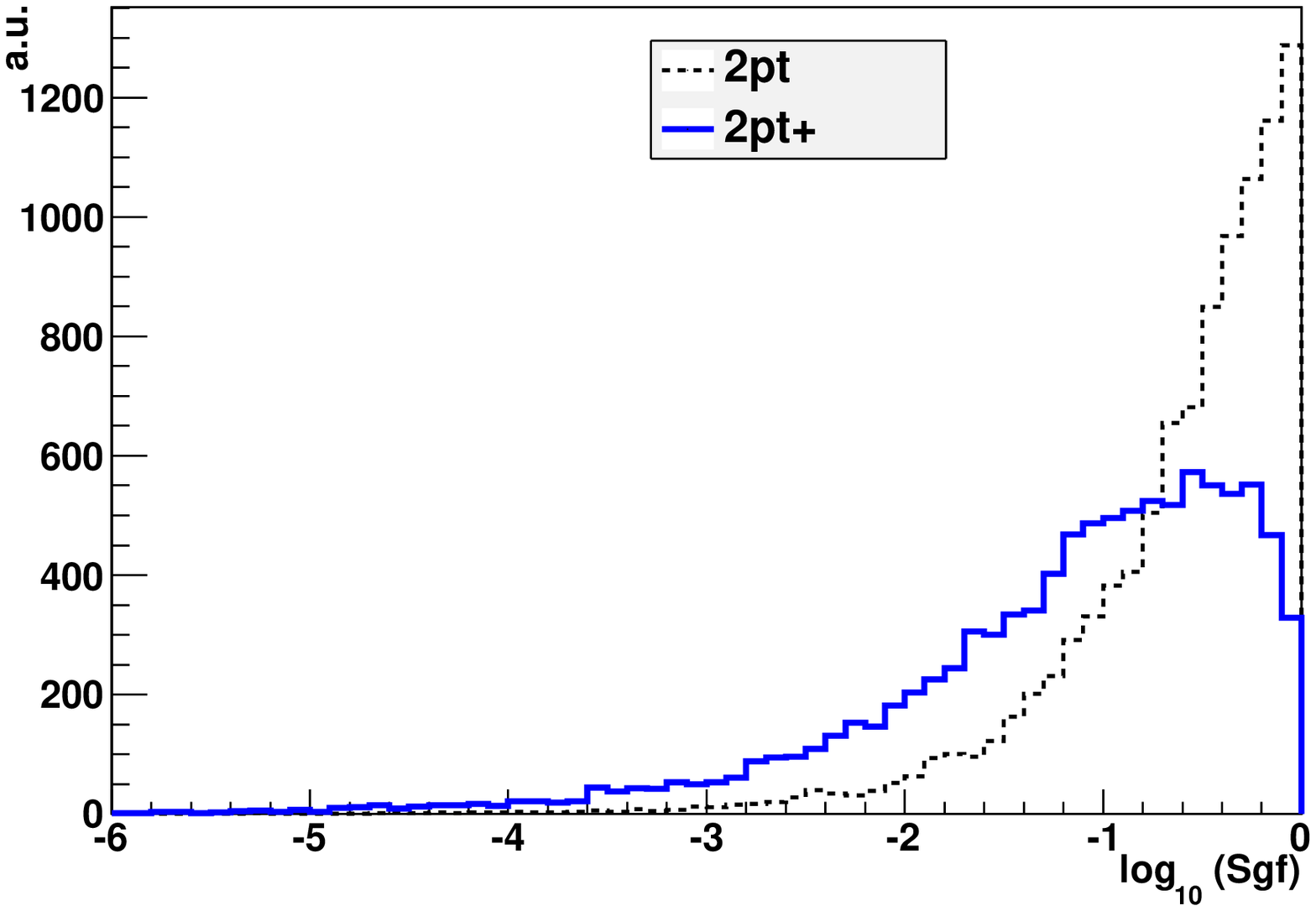}
\caption{Distribution of the significances of the 2pt+ and differential 2pt when 60 events are drawn from a VCV catalogue of sources with $z\leq 0.018$ and 3.2 deg
smearing angle, summed to 50\% isotropic component, and taking into account the Auger South exposure. }
\label{VCV_dist}
\end{center}
\end{figure}

Finally, in Fig. \ref{VCV_dist} we plot, for $N=60$ and for the signal corresponding to the lower panel of Fig. \ref{VCV_sens}, a histogram of the significance values returned for different realizations of the specific experiment by the 2pt (black line) and the 2pt+ (red line) tests: not only is the peak (most frequent) significance of the 2pt+ test translated to smaller values compared to the 2pt test, but also the very-low-value tails are systematically more populated.


\section{Testing a different hypothesis}

The test can be also used to reject a hypothesis other than isotropy. Quantitatively, any hypothesis can be expressed as a probability density function $\frac{dP(cos\theta,\phi)}{dcos\theta d\phi}$
where $\cos \theta$ and $\phi$ are the polar angles on the sphere. For a
sphere of unit radius and the isotropic ansatz the probability density function is:
 $\frac{dP(cos\theta,\phi)}{dcos\theta d\phi}=\frac{1}{4\pi}$. If the underlying probability density function is not flat,
correlations between $\cos \alpha$, $\cos \beta$ and $\gamma$ would appear, complicating
the implementation of the test. Our proposed approach to deal with non-flat distributions is to apply a 
 change of coordinates to a reference system for which  $\frac{dP(cos\theta,\phi)}{dcos\theta d\phi}=\frac{1}{4\pi}$. For any hypothesis, a point on the
 sphere $(cos\theta_0,\phi)$ is mapped into another position on the sphere with the following polar coordinates:
\begin{eqnarray}
 \cos \theta^\prime(\cos \theta_0,\phi_0) &=& \int_{-1}^{cos\theta_0} \frac{dP}{d \cos \theta d\phi}(\cos \theta,\phi_0) d \cos \theta  \nonumber \\
 \phi^\prime(\cos \theta_0,\phi_0) &=&\int_{0}^{\phi_0} \frac{dP}{d \cos \theta d\phi}(\cos \theta_0,\phi) d\phi  \nonumber \\
\label{MakeMeFlat}
 \end{eqnarray}
This transformation has the following nice property: if we have $N$ points on the sphere, we can draw $N$ parallels and meridians
 in such a way that each one passes through a point. This determines a grid on the sphere. The transformation proposed
 deforms the grid, changing the area of each cell, but introduces no caustics. This transformation effectively separates (brings closer together) the
 points in overdense (underdense) regions of the sphere.

This approach is simple and general, and overcomes the difficulties of dealing with each case in particular. We recommend
the use of this approach even when the sphere is not complete.

\section{Discussion}

We have presented a novel statistical test optimized for testing sparsely sampled datasets (with as few as 20 datapoints) for compatibility with an isotropic distribution on the sky. The test is an enhanced version of the classic 2pt test, incorporating all the information utilized in the 2pt test (the angular distances between events, or, equivalently, the {\em length} of the vector connecting event pairs), as well as additional, independent information about the {\em orientation} of each vector connecting event pairs. We call this new test the 2pt+ test. 

We propose here a specific method for the implementation of this test, based on distinct pseudo-likelihood estimators on a binned parameter space. The individual significances from each estimator are then combined using Fisher's method. Because of small correlations introduced by working in pair space rather than event space, the resulting quantity is not uniformly distributed as required for a true significance, so the final result we quote as significance is first corrected (its distribution is first rendered uniform) by Monte-Carlo simulations.  We additionally discuss a proposed methodology for utilizing the 2pt+ test to assess the compatibility of an event set with a distribution different from isotropy. 

We have tested the behavior of the test as a function of parameter space binning, number of events, and type of anisotropy. Based on these tests,  we recommend a binning choice of $\mu = 5$. The sensitivity of the test increases with increasing number of events, and, depending on the type of anisotropy, the method can saturate to significances better than $10^{-4}$ for as few as 60 events. 

We have also compared the sensitivity of the 2pt+ test with that of the 2pt test for different types of anisotropy and have found that the additional information encoded in the 2pt+ test results in yielding consistently better significances over the 2pt test. 

\section*{Acknowledgements}

We thank the Pierre Auger Collaboration for inspiring this work and for many discussions about testing isotropy with sparsely sample spherical datasets. This work was supported in part by NSF PHY-0758017 and by  the KICP under NSF PHY-0551142 at the University of Chicago. VP acknowledges support by NASA through the GLAST Fellowship Program, NASA Cooperative Agreement: NNG06DO90A. TV acknowledges support by the NSF Graduate Research Fellowship Program.

\end{document}